\documentstyle[12pt]{article}
\begin{document}
\title{On quantum black hole
entropy and Newton constant renormalization}
\author{ J. L. F.
Barb\'on,
\thanks{\tt barbon@puhep1.princeton.edu } \\ Joseph Henry
Laboratories, Princeton University,\\ Princeton, New Jersey 08544,
USA\\ and \\ R. Emparan,
\thanks{\tt wmbemgar@lg.ehu.es} \\
Departamento de F{\'\i}sica de la Materia Condensada,\\ Universidad
del Pa{\'\i}s Vasco, Apdo. 644,\\ 48080 Bilbao, Spain}
\date{}
\maketitle
\pagestyle{empty}
\thispagestyle{empty}
\begin{abstract}
We discuss the status of the black hole entropy formula $S_{{\rm BH}} =
A_H /4G$ in low energy effective field theory. The low energy expansion
of the black hole entropy is studied in a non-equilibrium situation: the
semiclassical decay of hot flat space by black hole nucleation. In this
context the entropy can be defined as an enhancement factor in the
semiclassical decay rate, which is dominated by a sphaleron-like saddle
point. We find that all perturbative divergences appearing in Euclidean
calculations  of the entropy can be renormalized in low energy
couplings.  We also discuss  some formal aspects of the relation
between the Euclidean and Hamiltonian approaches to the one loop
corrections to black hole entropy and geometric entropy, and we
emphasize the virtues of the use of covariant regularization
prescriptions.  In fact, the definition of black hole entropy in terms
of decay rates {\it requires} the use of covariant measures and
accordingly, covariant regularizations in path integrals. Finally, we
speculate on the possibility that low energy effective field theory
could be sufficient to understand the microscopic degrees of freedom
underlying black hole entropy. We propose a  qualitative physical
picture in which black hole entropy refers to a space of quasi-coherent
states of infalling matter, together with its gravitational field. We
stress that this scenario might provide a low energy explanation of
both the black hole entropy and the information puzzle.

\end{abstract}
\vfill
\begin{flushleft}
PUPT-1529, EHU-FT 95/5\\ hep-th/9502155\\ February 1995
\end{flushleft}
\newpage\pagestyle{plain}
\section{Introduction}\label{intro}
Black hole entropy has a neat
phenomenological meaning. During the late stages of the collapse
process in which a large black hole radiates thermally (i.e.,
according to Hawking's calculation \cite{haw}),  the interaction of
the black hole with the rest of the world occurs as if it had an
effective density of states $\rho\sim \exp (A_H/ 4G)$, where $A_H$ is
the horizon area and $G$ is the low energy (renormalized) Newton
constant. This is obtained from the Hawking temperature formula $T_H=
(8\pi GM)^{-1}$ and the equilibrium equation $\partial S_{{\rm
BH}}/\partial M=T_H^{-1}$.

A seemingly equivalent phenomenological derivation of black hole
entropy due to 't~Hooft \cite{tho}  does not assume any thermal
equilibrium boundary conditions (which are probably unphysical for
real collapsing black holes). Instead, this derivation is based on a
comparison of semiclassical absorption and emission rates and, in
contrast, the main assumptions are those of unitarity and {\it CPT}
invariance! The absorption cross section of the black hole would be
proportional to the horizon area:  \begin{equation}\label{absor}
\sigma_{\rm in}\sim |H_I ({\rm in})|^2 \rho_{{\rm BH}}(M)\sim A_H
\sim (GM)^2\,, \end{equation} where the interaction is described by a
Hamiltonian $H_I$ in some underlying quantum description. According
to Hawking, the emission rate for particles of mass $\delta M$ has a
thermal profile: \begin{eqnarray}\label{profile} \Gamma_{\rm
out}&\sim& |H_I ({\rm out})|^2 \rho_{{\rm BH}}(M-\delta M) \nonumber\\
&\sim& A_H e^{-\beta_H \delta M}\sim (GM)^2 e^{-8\pi GM\delta M}\,.
\end{eqnarray} Now, from {\it CPT} invariance and unitarity,
$|H_I({\rm in})|=|H_I({\rm out})|$, and one obtains the semiclassical
relation \begin{equation}\label{thoofts} {\rho_{{\rm BH}}(M-\delta
M)\over \rho_{{\rm BH}} (M)}\sim  e^{-8\pi GM\delta M}\,,
\end{equation} which is satisfied by $\rho_{{\rm BH}}(M)=e^{S_{{\rm
BH}}}=\exp (4\pi GM^2+C)$.

This phenomenological derivation relies on the Hawking emission
formula, which is a low-energy result. Hence, one would expect that a
microscopic description of the degrees of freedom leading to $S_{{\rm
BH}}$ could be achieved in low-energy effective field theory, at
least for large enough black holes, which accurately follow the
Hawking radiation formula. Strikingly, the only such calculation of
$S_{{\rm BH}}$ giving the correct result $A_H/4G$ regards the entropy
as a purely classical entity, without any statistical interpretation
\cite{gih}. This `intrinsic' entropy is at odds with the
phenomenological notion presented above. Several proposals for an {\it
ab initio} quantum construction of the Bekenstein-Hawking entropy have
been put forward over the years (for a review see \cite{bek}).
Notably, geometric (or entanglement) entropy has been proposed as
responsible for all or part of $S_{{\rm BH}}$ \cite{tho,frn}.

Entanglement entropy \cite{bkl,sre}  arises when the support of
physical operators is conventionally restricted to a proper region of
space, and finds its origin in quantum correlations across the
boundary between both regions. It is a fully quantum object naturally
scaling as the area, but it is ultraviolet divergent in field theory.
Moreover, this divergence poses subtle conceptual questions regarding
its physical interpretation. Perhaps one should operationally choose
a physical cutoff such that all the entropy of large black holes
comes from entanglement and its value is precisely
 $A_H/4G$, with $G$ the low-energy Newton constant.  Concrete
pictures of  this kind include estimates of the cutoff based on
horizon fluctuations \cite{frn}. Other suggestions involving
non-trivial quantum gravity physics are, for example, the hypothesis
of a `holographic' description of the state of collapsed matter
\cite{th2}, and the idea that black hole entropy looks classical
because it lives in a Hilbert space of states which cannot be realized
in field theory, such as special (perturbative) string configurations
\cite{suu} (the question of one-loop corrections in string theory is a
subtle one; see Refs.\ \cite{str}). A striking feature of all these
scenarios, in which one invokes subtle effects of quantum gravity to
target the value $A_H/4G$, is that automatically Hawking's
semiclassical calculation becomes suspect.  One could wonder that not
only black hole entropy, but also Hawking's temperature would arise as
`miraculous' successes of the semiclassical approximation.

A different alternative, which is compatible with some of the ideas
above, has been advocated by Susskind and Uglum in \cite{suu}.
According to this proposal geometric entropy is just a correction to a
classical entropy, and the divergences organize in such a way that
they renormalize the Newton constant, such that $S_{{\rm BH}} = A_{H}
/4G$ to all orders.  In this context, it is important to note that
the problem of understanding black hole entropy can be addressed
independently of the details of the endpoint of black hole
evaporation. Phenomenologically, one assigns entropy to a black hole
only during the period in which it is radiating thermally. For this
reason, a concrete low energy picture should be attainable to the
extent that Hawking's emission formula can be considered as accurate
at least for some period of time.

Many of the discussions of black hole entropy are carried out in the
Euclidean formalism for the Canonical Ensemble where, as stated
before, there is a classical contribution whose  explanation in
physical terms is not clear. In addition, the one-loop fluctuation
determinant in the gravitational sector has a negative eigenvalue
which leads formally to a complex one-loop entropy. While this  may be
sufficient to call into question the relevance of the whole approach,
it is likely that the imaginary part of the effective action still
has a physical meaning interpreted as a decay rate for black hole
nucleation. This is the interpretation of Gross, Perry and Yaffe in
\cite{gpy} of the original Hawking-Gibbons calculation. In the light
of these comments, it becomes interesting to recast the ideas of
\cite{suu} about entropy renormalization in the  language  of decay
rates, rather than in the static definitions of entropy.

There is an elegant operational definition of black hole entropy
appropriate for non-equilibrium situations, based on the fact that
semiclassical decay processes can be computed by means of Euclidean
instanton methods. For example, when charged black holes can be pair
created in a background electromagnetic field, the total rate may be
written as \begin{equation} \label{roeff} \Gamma \sim | {\rm
amplitude}|^2 \rho_{\rm eff}\,,   \end{equation} where  $\rho_{\rm
eff}$ is the effective density of final states. The classical
instanton contribution is such that  $\rho_{\rm eff}$ is enhanced for
non-extremal black holes with respect to the extremal ones (and to
monopole pair production)  by precisely the Bekenstein-Hawking
degeneracy factor $\rho_{{\rm BH}} = e^{A_H /4G}$ \cite{gsa}.

It was pointed out in Ref.\ \cite{gid} that the quantum corrections
(fluctuation determinant) to Eq.\ (\ref{roeff}) diverge uncontrollably
because the gaussian fluctuation determinant contains the factor
\begin{equation} \label{traza} {\rm Tr}_{{\cal H}_{ph}} e^{-\beta_H
H^{(0)}}\,, \end{equation} where $\beta_H$ is the Hawking inverse
temperature  (the black holes are created in equilibrium with the
Hawking radiation), and $H^{(0)}$ is the free Hamiltonian for the
physical (transverse) fluctuations around the instanton. This quantity
diverges because of the continuous spectrum of field excitations in
the black hole background \cite{tho}, in a way which makes the
corresponding renormalization a subtle question. The reason is that
the divergence can be traced back to the presence of the horizon as an
infinite red-shift surface. One may then argue that, after low energy
couplings have been renormalized according to,  e.g., graviton
scattering far from the black hole, the horizon is still in place and
the spectrum in a finite box is still continuous.  According to this
argument, it would appear that the calculation of the quantum
corrections to $S_{{\rm BH}}$ requires explicit knowledge of
Planckian physics and therefore the phenomenological
 formula $S_{{\rm BH}} = A_H /4G$ could not be substantiated  in a low
energy effective description. This state of affairs would contradict
the low energy theorem of Susskind and Uglum  \cite{suu}, which states
that an inambiguous low energy expansion exists for  $S_{{\rm BH}}$.
Whereas a breakdown of the effective  description could be expected
in a remnant scenario, one would  regard it as unreasonable if the
classical theory failed to provide an adequate description of large
enough black holes.

One of the aims of this paper is to show that a prescription can be
given so that such a breakdown does not occur. This will require
disentangling some of the subtleties involved in different ways of
considering the quantum black hole entropy. We choose to discuss
these issues in the context of thermal nucleation of black holes:
this provides a scenario rich in both technical features and
appealing interpretations in physical terms. The final outcome of our
analysis will be a low energy characterization of black hole entropy.

The paper is planned as follows. In the next section we cast the low
energy theorem of Ref.\ \cite{suu} into the language of Eq.\
(\ref{roeff})  and find that the continuous spectrum problem is
absent in the Euclidean formalism, where all ultraviolet divergences
can be renormalized in a standard way.  In Sec.~\ref{condis} we turn
to a more  detailed study of Eq.\ (\ref{traza}). We derive a
Euclidean prescription which is markedly different from the conical
singularity formalism, and makes manifest the problem of the
continuous spectrum. It  also explains why  this problem does not
appear in the treatment of Sec.~\ref{thermal}. Roughly speaking, what
happens is that the divergent Hamiltonian partition function Eq.\
(\ref{traza}) admits several formal representations as a determinant.
One of them uses an operator with continuous spectrum and the natural
regularization is non-covariant (a brick wall). The other
representation uses a fully covariant operator with discrete
spectrum, which is the one appearing in the calculation of decay rates
following the formalism of \cite{gpy}.

In Sec.~\ref{rsmirror} we offer some heuristic arguments showing that
the continuous spectrum problem is also absent in semiclassical
models for black hole collapse. It becomes an artifact of the eternal
black hole geometry as an asymptotic approximation, and the structure
of ultraviolet divergences should be again renormalizable in low
energy couplings in the standard way. Finally, in the last section we
speculate on a physical picture for the quantum origin of black hole
entropy. We point out that the space of quasi-coherent states of the
infalling matter and gravitational field could be used to parametrize
the microscopic degrees of freedom of black hole entropy. This point
of view does not necessarily rely on Planckian physics. One of the
appendices contains a calculation that would otherwise disrupt the
main line of the text. The other develops the subject of the thermal
instabilities of the vacuum of two-dimensional dilaton gravity.

\section{Thermal nucleation of black holes and entropy}
\label{thermal}

Black hole entropy as a classical enhancement factor of the final
state degeneracy may be studied in a technically  simple situation,
which nevertheless retains many physical features of the charged pair
production, at least for large black holes. This is the thermal
nucleation of neutral black holes in hot flat space, as studied in
Ref.\ \cite{gpy}, where it was computed to one loop order in the
dilute instanton gas approximation. The corresponding instanton is
simply the Euclidean section of the Schwarzschild geometry, which
mediates the nucleation of black holes of critical mass $M=\beta /8\pi
G$ inside a thermal bath of gravitons in flat space at temperature
$T=1/\beta$. The rate per unit volume is given by \cite{gpy}
\begin{equation} \label{nrate}    \Gamma = {\omega_0 \over 2\pi
\beta}\; {m_{P\ell}^3 \over 64 \pi^3}\;  (\mu\beta)^{212/45}\;
\exp\left(-{m_{P\ell}^2 \over 16\pi T^2}\right)\,. \end{equation}
 The relation between the imaginary part of the free energy and the
nucleation rate in this expression is the one corresponding to
semiclassical excitation over the barrier (see Ref.\ \cite{aff})
\begin{equation} \label{affleck} \Gamma = {\omega_0 \beta \over \pi}
{\rm Im}\left({F\over V} \right)\,. \end{equation} In other words,
the Euclidean instanton is better thought of as a sphaleron. Indeed,
it is time independent (the Schwarzschild metric is static) and the
fluctuation determinant in the physical sector has exactly one
negative eigenvalue $\lambda = -\omega_0^2 \simeq -0.19 /(GM)^2$,
responsible for the appearance of an imaginary part of the free
energy. The term proportional to $m_{P\ell}^3 $ in Eq.\ (\ref{nrate})
comes from the integration over the collective coordinates of the
sphaleron, and the term $(\mu\beta)^{212/45}$ appears because of the
anomalies associated with the Euler number counterterm, which is
non-vanishing in the Euclidean Schwarzschild section. The mass scale
$\mu$ appears as a dimensional transmutation (analogous to
$\Lambda_{\rm QCD}$) of the dimensionless Euler number coupling,
which then becomes a running coupling,  and may be phenomenologically
determined (a natural value in this context is $\mu\simeq m_{P\ell}$).
Finally, the exponential suppression factor comes entirely from the
classical gravitational action $e^{-I_{\rm cl}}$, which is given to
leading order by the Hilbert action \begin{equation} \label{hilbert}
I_H = -{1\over 16\pi G} \int_{\cal M} R -{1\over 8\pi G}
\oint_{\partial\cal M} (K-K^{(0)})\,. \end{equation}

The interpretation of Eq.\ (\ref{nrate})  according to the Fermi rule
Eq.\ (\ref{roeff}) is based on the fact that, due to the non-trivial
topology of the Euclidean Schwarzschild section the classical
suppression factor is not exactly the Boltzmann factor $e^{-\beta
M}$.  Indeed, on a manifold with cylindrical topology (a usual
thermal manifold), the Hilbert action equals the canonical action and
$I_{\rm cl} ({\rm cylinder} \times S^2) = \beta M_{\rm  ADM}$. On the
other hand, on the Schwarzschild manifold, with topology ${\cal M} =
{\rm Disk}\times S^2$ there is one boundary missing, which produces
$I_{\rm  cl}({\cal M})  = {1\over 2}\beta M_{\rm ADM}$. The Boltzmann
factor is in excess exactly by the value of the classical black hole
entropy (recall $\beta=8\pi G M$ for the nucleated black holes):
\begin{equation} \label{enhan} e^{-{\beta^2 / 16\pi G}} = e^{-\beta
M}\; e^{4\pi G M^2}= e^{-\beta M}\rho_{{\rm BH}}\,. \end{equation}

Hence, an operational definition of black hole entropy in this context
would be the excess of the classical action over the Boltzmann factor
\begin{equation} \label{entdef} S_{{\rm BH}} \equiv \beta M - I_{\rm
cl}({\cal M})\,.  \end{equation} Since the gravitational sector may
be regarded as a low energy effective theory, quantum corrections
require the introduction  in $I_{\rm cl}$ of the whole tower of
counterterms, leading to a low energy expansion \begin{equation}
\label{expansion} S_{{\rm BH}} = {A_H \over 4G} + \sum_{n} {\lambda_n
\over m_{P\ell}^{d_n -4} } \int_{\cal M} {\cal O}_n\,. \end{equation}
Here we have suppressed the cosmological constant counterterm, since
asymptotic flatness is a condition of the problem. The leading term
absorbs the renormalization of the Newton constant, while the others
are also phenomenologically determined. It is important to note that
the  definition of $S_{{\rm BH}}$ given in Eq.\ (\ref{entdef})  is
not in general equivalent to others based on the thermodynamic formula
\begin{equation} \label{termo} S = (\beta \partial_{\beta} -1) I_{\rm
cl}\,.     \end{equation} In the original analysis of Gibbons and
Hawking \cite{gih},  the derivative was taken on the space of
classical solutions of the field equation. Alternatively, in the
conical  singularity method \cite{btz,sus}, one holds $M$ fixed while
varying $\beta$, thus going  off-shell due to the conical singularity
at the horizon.

Actually, one may regard the sphaleron-like interpretation of the
Euclidean Schwarzschild instanton \cite{gpy} as more natural than the
thermostatic interpretation behind formula (\ref{termo}), because the
negative mode at one loop signals an instability of the canonical
ensemble for thermal gravitons, in addition to the infrared Jeans
instability. Yet, the low energy effective theory predicts a value for
the decay rate which may have physical meaning. The definitions based
on Eqs.\ (\ref{entdef}) and (\ref{termo}) will give, in general, a
slightly different low energy expansion, although the leading term
seems to be universal (in a fashion similar to the first two terms of
the beta function in gauge theories, which are independent of the
definition of physical coupling).

A very striking feature of Eq.\ (\ref{nrate}) is the absence of the
potentially troublesome partition function in Eq.\ (\ref{traza}). In
fact, it is easy to see that it cancels against the flat space
normalization, up to an ultraviolet finite boundary or `surface
tension' term. To be more specific, let us consider the total
partition function given as a dilute multi-instanton sum:
\begin{equation}\label{multiz} Z=\exp(-\beta{\rm Re}
F+{i\pi\over\omega_0}V\Gamma)=\sum_{N=1}^{\infty} {1\over N!}Z_{(N)}
e^{-I_{\rm cl}^{(N)}}\,, \end{equation} where $I_{\rm cl}^{(N)}\simeq
NI_{\rm cl}^{(1)}=N\beta M/2$ and $Z_{(N)}$ denotes the perturbative
partition function around the $N$-multiblackhole solution. To one
loop order one finds \begin{equation}\label{zetan} Z_{(N)}=({i/ 2})^N
C^N {\rm det}_+^{-1/2}(I_{(N)}'')\,, \end{equation} where $C$ stands
for the contribution of the collective coordinates (zero modes) and
anomalous scaling. The factor $i/2$ comes from the usual half-contour
rotation for the $N$ negative modes and $I_{(N)}''$ is a combination
of second order elliptic differential operators which includes
fluctuation kernels for the physical as well as unphysical graviton
polarizations, and the corresponding ghost terms \cite{ghp}. Roughly
speaking, the ghost determinant cancels the longitudinal and trace
parts of the graviton excitations, leaving the physical
(transverse-traceless) fluctuations.

In any covariant regularization, the ultraviolet divergences in the
perturbative effective action $W_{\rm eff}={1/2}\log{\rm
det}(I_{(N)}'')$ can be absorbed in the counterterm series of $I_{\rm
cl}$. A very convenient one-loop prescription is given by $\zeta$
function regularization, which only requires the spectrum of $I''$ to
be discrete. This is always the case at finite volume, since  all
operators are elliptic and the Euclidean manifold is compact with
boundary. If we write the ultraviolet finite part of $W_{\rm eff}$ as
a volume integral it is clear that, within the dilute gas
approximation and in the large volume limit, $W_{\rm eff}$ is
dominated by the free energy of gravitons in flat space. Let us
separate the contribution of the asymptotic thermal gravitons from
those close to the horizon (up to, say, a radius $r\sim 3GM$). Then
one finds \begin{equation}\label{seff} W_{\rm eff}\simeq N W_{\rm
hor}+\beta f_g (V-NV_{{\rm BH}}) \,, \end{equation} where
$f_g=-\pi^2/45\beta^4$ denotes the free energy density of gravitons
in flat space, and $V_{{\rm BH}}$ is the excluded volume per black
hole. If we multiply and divide by the flat space partition function
$Z_{(0)}=\exp (V\pi^2/45\beta^3)$ we get an overall factor in the
$N$-instanton term  \begin{equation}\label{bound} Z_{(0)} e^{-N\beta
F_B}\,, \end{equation}  where $F_B$ is a boundary free energy given
by the contribution to $W_{\rm eff}$ coming from the horizon region
minus the graviton free energy in the same volume of flat space
\begin{equation}\label{boundf} \beta F_B=W_{\rm hor}-\beta f_g
V_{{\rm BH}}\,. \end{equation} Notice that $\beta f_g V_{{\rm BH}}$
is a pure number, independent of $M$. In fact, $\beta F_B$ appears as
a constant term in the $1/V$ expansion of $W_{\rm eff}$:
\begin{equation}\label{weff} W_{\rm eff}=\beta f_g V+N\beta F_B+
O(V^{-1})\,. \end{equation} For small black holes (corresponding to
high temperature) this term should approach zero, whereas for large
black holes (i.e., low temperature), $W_{\rm hor}$ scales like the
vacuum energy of Euclidean Rindler space, $W_{\rm hor}\sim
(M/m_{P\ell})^4 +{\rm const.}$. Hence one concludes that
\begin{equation}\label{freebound} \beta F_B\sim \left({M\over
m_{P\ell}}\right)^4 \bigl[1+O\bigl( {m_{P\ell}\over M} \bigr)\bigr]\,.
\end{equation} When the multi-instanton sum is performed the term
$\beta F_B$  exponentiates and it contributes to the imaginary part of
the free energy. The real part is given by the flat space free energy
as it should be, and the corrected low energy expansion for the rate
reads \begin{equation}\label{corrgamma} \Gamma_{\rm
corr}={\omega_0\over 2\pi\beta}\; {m_{P\ell}^3\over 64 \pi^3}\;
f(\mu\beta)e^{-\beta M}e^{S_{{\rm BH}} } e^{-\beta F_B}\,,
\end{equation} where $S_{{\rm BH}}$ is defined in Eq.\
(\ref{expansion}) and $f(\mu\beta)=(\mu\beta)^{212/45}+\dots$, is
obtained from the perturbative expansion of the $\beta$ function
associated to the dimensionless coupling to the Euler number.

The boundary term $\exp(-\beta F_B)\simeq\exp[-C(M/m_{P\ell})^4]$
dominates the suppression factor at low temperatures. This agrees with
the fact that surface effects become increasingly important: in the
large $M$ limit the Rindler region covers all of (Euclidean)  space.
The boundary partition function is thus related to the part of Eq.\
(\ref{traza}) coming from the vicinity of the horizon.

In the construction presented above, black hole entropy is
fundamentally a classical object with no microscopic interpretation,
and quantum corrections organize in a low energy expansion.
Furthermore, the renormalization of the Newton constant implied in the
definition of $S_{{\rm BH}}$ is the same that one would obtain from
graviton scattering far from the black hole, as long as a covariant
procedure, such as $\zeta$-function regularization, is employed
everywhere. This, in turn, is ensured by the fact that the finite
volume Euclidean manifold is compact and smooth and  at the
equilibrium temperature there is no global distinction between finite
temperature free energy and vacuum energy.

 An important point to stress is that, at least for non-extremal black
holes, the problem of continuous spectrum is absent from the previous
discussion. All the operators involved are elliptic, and have discrete
spectrum at finite volume. An explicitly covariant regularization is
possible and there is no obstruction to the low energy theorem of
\cite{suu}. For example, having discrete spectrum one can use $\zeta$
function regularization at one loop, in which there are no
divergences at all and the total black hole entropy  comes out
clearly as $A_H /4G$.

\section{Continuous  versus discrete fluctuation spectrum}
\label{condis}

In this section we study some aspects of the Hamiltonian partition
function, Eq.\ (\ref{traza}),  which following Unruh \cite{unr}, is
related to the entanglement density matrix in the vacuum of the
extended eternal black hole geometry. We start by reviewing the
disease caused by continuous black hole spectrum, as first pointed
out by 't~Hooft in Ref.\ \cite{tho}, and work backwards to derive an
Euclidean formulation which makes manifest the differences between
Eq.\ (\ref{traza}) and the term $\exp(-\beta F_B)$ that we have found
in the previous section.

\subsection{Statistical mechanics of the fluctuation degrees of
freedom}
Although we have in mind the physical situation studied in
section \ref{thermal} (thermal gravitons) the discussion may be
generalized to different matter contents. In general, let the
gaussian  (Lorentzian) action for quadratic fluctuations around the
black hole be \begin{equation}\label{quadfluct} S^{(2)}={1\over
2}\int_{\cal M} \varphi\;{\cal L}\;\varphi\,, \end{equation} where
${\cal L}=-\nabla^2+V(g)$ and $\varphi$ represents the physical
(transverse) excitations. For example, for a scalar field we have
${\cal L}=-\nabla^2+m^2+\xi R+\dots$, while for transverse-traceless
gravitons---the case relevant to the previous section---we must
consider ${\cal L}_{TT} h_{\alpha\beta}=-\nabla^2
h_{\alpha\beta}-2R_{\alpha\beta \gamma\delta} h^{\gamma\delta}$ (we
are focusing on bosonic fields for simplicity).

Choosing a time slicing adapted to the Killing vector
$\partial/\partial t$, where $t$ is the asymptotically minkowskian
time, we may express the free canonical Hamiltonian associated to
Eq.\ (\ref{quadfluct}) in terms of the eigenfrequencies as
\begin{equation}\label{Hamilt} H^{(0)}=\sum_\omega \omega
a^{\dagger}_\omega a_\omega  +\Lambda_B\,, \end{equation} where
$\Lambda_B={1\over 2}\sum_\omega \omega$ is the (Boulware) vacuum
energy, formally infinite, and $a^{\dagger}_\omega$ generate the
physical Fock space. The one loop free energy takes then the well
known form \begin{equation}\label{freef} \beta
F-\beta\Lambda_B=\sum_\omega \log(1-e^{-\beta\omega})\,.
\end{equation} At finite volume, this quantity is ill defined even
with an ultraviolet cutoff in the frequency sum: $\omega\leq\Lambda$.
The reason is that the spectrum of a black hole in a box is still
continuous because the horizon behaves as a non-compact boundary. The
eigenvalue problem for the frequencies is
\begin{equation}\label{eigenfreq} (-g_{00})(-\vec\nabla^2
+V(g))\psi_\omega({\bf x}) =\omega^2 \psi_\omega({\bf x})
\end{equation} In tortoise coordinates $r_*=r+2GM \log|r/2GM-1|$this
is a Schr{\"o}dinger problem for radial excitations with $L^2$
metric, and with an effective potential  $V_{\rm eff}\propto
-g_{00}\sim \exp (4\pi T_H r_*)$  as we approach the horizon
($r_*\rightarrow -\infty$). As a result, the spectrum is continuous
unless a horizon regulator (brick wall) is imposed. This all looks
very different from the discussion in the preceeding section, where
all operators would present discrete spectrum after standard infrared
regularization (large volume cutoff).

In particular, as pointed out in Ref.\ \cite{gid}, the problem of
continuous spectrum seems to remain even after the Newton constant has
been renormalized according to graviton scattering far from the black
hole, because it only depends on the existence of the horizon as an
infinite red-shift surface. Heuristically, the brick wall boundary
condition is a local ultraviolet cutoff, because the condition
$\omega\leq\Lambda$ is not a uniform cutoff for local static
observers, who measure local frequencies $\omega_{\rm loc}
=\omega/\sqrt{-g_{00}}$. Thus, the brick wall cuts off unphysical
static frames. It is, however, very disturbing that this
interpretation of the cutoff is frame dependent.  This is a first
indication of the fact that the continuous spectrum  can not be
easily cut-off in a covariant way.

In order to bring the discussion to the terms of  section
\ref{thermal}, it is necessary to rewrite the free energy Eq.\
(\ref{freef}) in Euclidean form. This can be done directly, as in
flat space, by means of the $\zeta$ function identity (we follow
Ref.\ \cite{bar}) \begin{equation}\label{zetafun} \prod_{n=1}^\infty
(A+n^2/B)= {2\over\sqrt{A}}\sinh  (\pi\sqrt{AB})\,. \end{equation} We
get \begin{equation}\label{logdet} \beta F={1\over 2} \log
\prod_{n\in Z}\prod_{\omega} (4\pi^2n^2/\beta^2 +\omega^2)\equiv
-\log{\rm det}^{-1/2}(\widehat{\cal L})\,. \end{equation} This
defines $\widehat{\cal L}$ as the operator with eigenvalues $4\pi^2
n^2/\beta^2 +\omega^2$. From Eq.\ (\ref{eigenfreq}) we conclude that
$\widehat{\cal L}$ is given by  \begin{equation}\label{hatl}
\widehat{\cal L} = -\partial_\theta^2 + (-g_{00}) (-\vec\nabla^2+V(g))
\end{equation} acting on periodic functions of the Euclidean time
$\theta$ of the form $\hat\psi_{n,\omega}=e^{2\pi i n\theta/\beta}
\psi_\omega ({\bf x})$, where $\psi_\omega({\bf x})$ are the spatial
harmonics in Eq.\ (\ref{eigenfreq}).

Curiously enough, this is not the covariant fluctuation operator, but
rather a local multiple: \begin{equation}\label{eles} \widehat{\cal
L}=(-g_{00}){\cal L}\,. \end{equation} The inner product for
$\widehat{\cal L}$, as inherited from the $L^2$ inner product in
tortoise coordinates (or the Klein-Gordon metric in Lorentzian
signature) is \begin{equation}\label{kgprod}
\langle\hat\psi|\hat\psi' \rangle = \int_{\cal M} d^4 x
{\sqrt{-g}\over (-g_{00})}\; \hat\psi^* \hat\psi'\,. \end{equation}
{}From this we can derive a path integral formula. In general, given
an inner product   \begin{equation}\label{inprod}  \langle
\psi,\psi'\rangle_\rho = \int d^4 x\; \rho(x) \psi^* (x) \psi' (x)
\,,     \end{equation} the determinant of an operator ${\cal L}_\rho$
admits the representation \begin{equation}\label{detl} {\rm
det}^{-1/2}({\cal L}_\rho) = \int {\cal D}_\rho \varphi
\exp\biggl(-{1\over 2}\langle\varphi, {\cal L}_\rho
\varphi\rangle\biggr)\,, \end{equation} where the measure is formally
given by \begin{equation}\label{measure} {\cal D}_\rho \varphi=
\prod_{n}{d c_n\over\sqrt{2\pi}}  =\prod_{x}
{d\varphi_x\over\sqrt{2\pi}} \rho_x^{1/2}  \,. \end{equation} Here
$c_n$ are the Fourier coefficients of the field $\varphi$ in a basis
orthonormal with respect to the product  (\ref{inprod}).

It is interesting that the inner product Eq.\ (\ref{kgprod}) precisely
gives the action $S^{(2)}$ in the exponent,
\begin{equation}\label{action} S^{(2)}={1\over 2}\int d^4 x
\sqrt{-g}\; \varphi{\cal L} \varphi = \langle\varphi|\widehat{\cal
L}|\varphi\rangle\,. \end{equation} So, the operator $\widehat{\cal
L}$ with the inner product Eq.\ (\ref{kgprod}) is {\it classically}
equivalent to the operator ${\cal L}$ with the covariant inner
product. However, quantum mechanically, there is a difference in the
path integral measure.

We have then established \begin{eqnarray}\label{result}  {\rm
Tr}_{{\cal H}_{ph}} e^{-\beta H^{(0)}}&=&{\rm det}^{-1/2}(
\widehat{\cal L}) \nonumber\\ &=&\int \prod_x {d\varphi_x\over
\sqrt{2\pi}}  \left({\sqrt{-g}\over -g_{00}}\right)^{1/2}_x
  e^{-S^{(2)}[\varphi]}\,. \end{eqnarray}   This result was also
obtained in Ref.\ \cite{dea} using the canonical derivation of the
path integral. It is remarkable because it shows that the canonical
partition function Eq.\ (\ref{traza}) is {\it not}  formally equal to
${\rm det}^{-1/2}({\cal L})$. Rather, it equals the determinant of a
related operator which is singular at the horizon where $g_{00}=0$.
Accordingly, the operator $\widehat{\cal L}$ has continuous spectrum
at finite volume, and does not admit $\zeta$ function regularization
unless we provide some kind of brick wall cutoff. Therefore, the
topology of the Euclidean manifold appropriate to $\widehat{\cal L}$
is cylindrical: $\widehat {\cal M}={\cal M}-\{ {\rm Horizon}\}$ is
non-compact in the vicinity of the horizon. If we would use this as
the physical thermal manifold, the classical contribution to the
entropy would vanish.

The peculiar topology associated to Eq.\ (\ref{result}) can be traced
back to its origin as geometric or entanglement entropy, at least when
it is calculated as a thermal sum. For example, if we consider the
entanglement entropy generated by performing a trace over half of
Minkowski space \cite{bkl}, the formal procedure to expose the
thermal nature of the density matrix uses a trick due to Unruh
\cite{unr} (see also Refs.\ \cite{laf,caw}).

One decomposes the total Cauchy surface into two non-compact left and
right components by an appropriate coordinate mapping, which in this
case is equivalent to the Rindler acceleration. Since the two
components are non-compact, in fact the origin (the position of the
boundary) is not part of the mapping. In other words, one writes
${\cal H}'={\cal H}_L\otimes{\cal H}_R$, where ${\cal H}_{L,R}$ are
the left and right Hilbert spaces, and ${\cal H}'$ is the total
Hilbert space minus the field oscillator at the boundary. This is the
formal origin of the missing point in the Euclidean manifold $\widehat
{\cal M}$.

The density matrix for the vacuum obtained by tracing out degrees of
freedom in the left half-space can be found to be \cite{laf,bpz,law}
\begin{eqnarray}\label{densmat} \langle\varphi|\rho|\varphi'\rangle
&=&\langle\varphi|({\rm Tr}_{{\cal H}_L}|0\rangle \langle 0|
)|\varphi'\rangle\nonumber\\ &=&\prod_\omega \left({\omega\over \sinh
2\pi\omega}\right)^{1/2} \exp \left(-{\omega\over
2}\left\{\coth 2\pi\omega(\varphi^2_\omega +{\varphi'}^2_\omega)
-{2\varphi_\omega{\varphi'}_\omega\over\sinh
2\pi\omega}\right\}\right)\,, \end{eqnarray} where $\varphi_\omega$,
${\varphi'}_\omega$ are the Fourier components of the spatial fields
in the right half-space,  analyzed in the basis of spatial
eigenfunctions $\psi_\omega({\bf x})$ orthonormal with respect to the
spatial section of the inner product (\ref{kgprod}).  The exponential
term in Eq.\ (\ref{densmat}) corresponds to the classical action
$S^{(2)}$ between configurations $\varphi, \varphi'$, whereas the
prefactor comes from the fluctuation determinant around the classical
path.  It is easy to check that to obtain it from the four
dimensional  Euclidean path integral one must use the noncovariant
measure in Eq.\ (\ref{result}), and  introduce $\varphi$, $\varphi'$
as the values of the field at each side of the cut along $\theta=0$
\cite{laf,caw}.

These results may seem disturbing at first, because they indicate that
the Euclidean construction for entanglement entropy is formally
defined in terms of $\widehat{\cal L}$ instead of the covariant
operator ${\cal L}$. On the other hand, we know that the
Hartle-Hawking Green's function defined without boundary condition on
the Euclidean section ${\cal M}$ is the correct thermal Green's
function for static observers. In fact, both Green's functions are
equal: $\widehat G(x,x')= G_{HH}(x,x')$ and there is no contradiction.
 Again, this follows easily from the freedom to choose different
operators provided the inner product is changed accordingly. The
Green's function of an operator ${\cal L}_\rho$ defined as
\begin{equation}\label{greenf} G_\rho (x,x') = \langle x|{\cal
L}_\rho^{-1}|x'\rangle \end{equation} satisfies the equation
\begin{equation}\label{greeneq} {\cal L}_\rho (x) G_\rho(x,x') =
\delta_\rho (x,x')= {\delta (x-x')\over \rho_x}\,. \end{equation}

Using the expression for $\widehat{\cal L}$ and $\hat\rho$ in terms of
${\cal L}$ and $\rho$ it is trivial to realize that $G_{HH}$ and
$\widehat G$ satisfy the same equation
\begin{equation}\label{greeneq2} {\cal L}_x \widehat G (x,x') =
{\delta (x-x')\over \sqrt{-g_x}}\,. \end{equation} Therefore,
$G_{HH}$ and $\widehat G$ are obviously identical  when the boundary
conditions are the same, such as in a brick wall model. For the no
boundary case, the equality is not obvious, because $\widehat
G(x,x')$ cannot be extended to the horizon in terms of the
eigenfunctions of $\widehat{\cal L}$. However, an explicit computation
in Rindler space can be done (see Appendix \ref{ap}) which ensures
$G_{HH}=\widehat G$ also in the no boundary case.

Thus, for local physics, the difference between $\widehat{\cal M}$ and
${\cal M}$ is just the way in which the no-boundary condition of
Hartle and Hawking is introduced. However, the difference is important
for the issue of the total number of states of the black hole in low
energy field theory.     This is due do the fact that, in going from
the Green's function to the extensive free energy, one has to give
sense to the expression \begin{equation}\label{trlog} {1\over 2} {\rm
Tr}_{\{ x\}}\log G(x,x)\,. \end{equation} Different prescriptions for
the spatial trace and the coincidence limit turn into the different
determinants above. The disease of continuous black hole spectrum
arises when one works with $\widehat G$, which leads to considering
${\rm det}^{-1/2}(\widehat{\cal L})$.  As explained above, this is
naturally regularized by means of a brick wall cutoff. On the other
hand, use of $G_{HH}$ in Eq.\ (\ref{trlog}) is concomitant to the
computation of ${\rm det}^{-1/2}({\cal L})$, which is free of the
continuous spectrum problem. In this case the regularization procedure
is fully covariant and we obtain the results of section \ref{thermal}.
In fact, both prescriptions are formally related by a conformal
transformation.        To see this, we recall that,  according to
Eq.\ (\ref{logdet}), there are many path integral versions of the
same determinant, because we can change the operator at the price of
rescaling the inner product (thereby changing the functional
measure). A particularly nice variation is given by the  `optical'
inner product, which is covariant with respect to the conformally
related (`optical')  metric $\bar g_{\alpha\beta}=g_{\alpha\beta} /
(-g_{00})$ and has weight $\bar\rho=\sqrt{-g}/(-g_{00})^{d/2}=({-\bar
g})^{-1/2}$ in $d$ spacetime dimensions. Then the operator
\begin{equation}\label{lopt} {\bar{\cal L}}\equiv (-g_{00})^{d+2\over
4}\;{\cal L}\;(-g_{00})^{ 2-d\over 4} \end{equation} has the same
eigenvalues as $\widehat{\cal L}$ and ${\rm det}^{-1/2}(\widehat
{\cal L})={\rm det}^{-1/2} (\bar {\cal L})$.

In the conformally invariant case, a nice relation between the
determinants of ${\cal L}$ and $\widehat{\cal L}$ can be written using
the optical operator as an intermediate step. A conformally invariant
fluctuation operator for scalars is given by
\begin{equation}\label{confl} {\cal L}_c (g) = -\nabla^2 + {d-2\over
4(d-1)} R\,. \end{equation} A simple computation shows that
$\bar{\cal L}_c={\cal L}_c(\bar g)$ and we may write
\begin{eqnarray}\label{dets} {\rm det}^{-1/2}(\bar{\cal L}_c)&=& {\rm
det}^{-1/2} ({\cal L}_c( \bar g))\nonumber\\ &=&\int\prod_x
{d\varphi_x\over\sqrt{2\pi}} (-\bar g_x)^{1/4}\nonumber\\&\times&
\exp{-\left[{1\over 2}\int\sqrt{-\bar g} \varphi{\cal L}_c(\bar
g)\varphi\right]}\,. \end{eqnarray} But the last path integral is
conformally related to the covariantly regularized path integral for
the operator in the physical metric.  Then we obtain
\begin{equation}\label{detsl} {\rm det}^{-1/2}(\widehat{\cal
L}_c)={\rm det}^{-1/2}(\bar{\cal L}_c) =e^{-I_L [\log g_{00}]} {\rm
det}^{-1/2}({\cal L}_c) \end{equation} (see also Ref.\ \cite{dea}).
In two dimensions $I_L$ is the standard Liouville functional,   while
in four dimensions it is in general a non-local action \cite{bod}.

\subsection{Brick wall regularization and renormalization}

Equations (\ref{dets},\ref{detsl}) only make sense with a brick wall
in place, because otherwise the non-compact operators have no
well-defined determinant.  In such a situation $I_L\sim \beta\times
{\rm finite}$, i.e., it contributes only to the vacuum energy (not to
the entropy). This means that one can compute  the entropy in the
presence of the brick wall directly from the ultraviolet finite part
of ${\rm det}^{-1/2} ({\cal L})$.  It is important to recognize that,
in the absence of a brick wall Eq.\ (\ref{detsl}) has a formal
status, because it relates infinite quantities.

The leading brick wall divergence is in fact independent of the
particular potential term occurring in ${\cal L}= -\nabla^2 +V(g)$,
provided $V(g)$ is regular in the horizon region. This is due to the
fact that the leading divergence depends only on the effective
potential $-g_{00} V(g)$, which vanishes exponentially in the horizon
region. The potentially troublesome angular degrees of freedom
\cite{kas}, which may spoil the accuracy of the WKB approximation,
sum up such that the WKB result is surprisingly correct (this is easy
to check by using Eq.\ (\ref{detsl}) and computing in the optical
metric \cite{emp,dea}).

The final answer for the leading divergence per degree of freedom in
$d>2$ dimension is \begin{eqnarray}\label{leadingdiv} S_{\rm
div}&=&-(\beta_H F-\beta_H \Lambda_B)_{\rm div}d \nonumber\\ &=&
 {d\; \Gamma(d/2)\;\zeta(d)\over (d-2) \pi^{3d/2-1} 2^{d-1}}{A_H\over
\epsilon^{d-2}_{\rm bw}}\,, \end{eqnarray} where $\epsilon_{\rm bw}$
is the brick wall cutoff.  In two dimensions $S_{\rm div}= 1/6
\log\epsilon_{\rm bw}^{-1}$. Also, for fermions one obtains the usual
statistical correction factor $S_{\rm Fermi}=(1-2^{1-d})S_{\rm
Bose}$.

At this point one can adopt different attitudes. If black hole entropy
is primarily regarded as a quantum object and Eq.\ (\ref{leadingdiv})
considered at least part of it, then the entropy is clearly cutoff
dependent. We cannot  predict its value using low energy quantum
gravity  nor understand what degrees of freedom $S_{{\rm BH}}$
accounts for. In this view, the final result $S_{{\rm BH}}=A_H/4G$
with $G$ the long distance Newton constant,  would seem to come out
in a rather `miraculous' way from Planckian dynamics in quantum
gravity. Variations of this idea have been put forward by various
authors \cite{tho,frn}.

Another possibility is to consider a classical entropy,
 and take Eq.\ (\ref{leadingdiv}) as a counterterm renormalizing
Newton constant.  However, there is some arbitrariness here since the
renormalization conventions appropriate for graviton scattering far
from the black hole and for Eq.\ (\ref{leadingdiv}) do not agree in
general. For example, the counterterms induced by a scalar field on
the vacuum energy are (in Schwinger proper time regularization)
readily found from the heat kernel expansion
\begin{eqnarray}\label{heat} \Lambda_{\rm counter}&=&-{{\rm
Vol}({\cal M})\over d
 (4\pi)^{d/2}}{1\over \epsilon^d}+ {\sqrt\pi {\rm Vol} (\partial{\cal
M})\over 2 (d-1) (4\pi)^{d/2}} {1\over \epsilon^{d-1}}\nonumber\\ &-&
{\epsilon^{2-d}\over (d-2)(4\pi)^{d/2}} \left( {1\over 6}\int_{\cal
M} R+{1\over 3}\oint_{\partial{\cal M}} K\right)+\dots \end{eqnarray}
The last term induces a renormalization (in four dimensions)
\begin{equation}\label{renorm} G^{-1}_{\rm bare}\rightarrow
G^{-1}_{\rm bare}+{1\over  12\pi\epsilon^2}\,. \end{equation}

Now, in order to compare Eqs.\ (\ref{leadingdiv}) and (\ref{heat})  we
would need an invariant relation between     both cutoffs. It is
unlikely that such a relation exists because, as we pointed out
before, the physical interpretation of the brick wall as an
ultraviolet cutoff is fundamentally frame-dependent.  If one insisted
on comparing the results, the only possible ``natural" relation
should be based on  the fact that    the Schwinger proper time cutoff
is a length cutoff for paths in the first quantized path integral
representation of determinants. Then,   one could declare that the
Schwinger cutoff is  set by the minimum length non-contractible path
in the brick wall manifold  \begin{equation}\label{eps}
\epsilon\simeq \epsilon_{\rm bw}{2\pi\beta\over \beta_H}\,,
\end{equation} and this  would lead to $S_{\rm
div}=(\pi/90)(A_H/\epsilon^2)$ in four dimensions, which could not be
absorbed with the renormalization above. As a consequence, if we
wanted the renormalization to work along the lines of section
\ref{thermal}, we would be led to {\it ad hoc} choices of brick wall
cutoff.

This situation may be summarized by saying that the use of brick wall
regulators has a heuristic value but,
 if we assume that there is an inambiguous classical entropy,  a
systematic treatment of the renormalizaton   procedure in low energy
theory requires the use of   covariant schemes,
 based on the Hartle-Hawking regular manifold ${\cal M}$---as in
section \ref{thermal}---, or a conical deformation of it (this agrees
with remarks made in recent papers \cite{rec}).  In fact, in the
context of the black hole nucleation approach, we can say that the
covariant method is the only possible choice.   This is due to the
fact that the continuous spectrum operator ${\widehat {\cal L}}$ has
positive spectrum by construction. There is no way we could get a
negative eigenvalue from this operator and thus no imaginary part for
the free energy. As a result, this operator cannot appear if we want
to maintain the physical picture of hot space decay.

\section{Redshift arguments in mirror models} \label{rsmirror}

In this section we argue on a heuristic basis that, in semiclassical
collapse models, the continuous spectrum problem seems to be spurious.
In  the WKB approximation one basically gets the results of naive
red-shift calculations, i.e., formula (\ref{leadingdiv}) can be
obtained from \cite{bar}  \begin{equation}\label{rshft} S_{\rm
div}\simeq {d\; \Gamma (d/2)\; \zeta(d) \over \pi^{d/2}} \int d({\rm
Vol}) (\beta_H\sqrt{-g_{00}})^{1-d}\,. \end{equation}

This suggests that the divergence in Eq.\ (\ref{leadingdiv})  should
be properly related to the unphysical observers close to the horizon.
Any quantity computed from $\widehat{\cal L}$ refers to a family of
static observers which become singular at the horizon---a physical
static frame at the horizon has infinite energy. Yet, this is an
artifact of the eternal black hole geometry as an effective
approximation to a collapse solution.  This point deserves further
explanation.

Hawking radiation is dynamically generated by the time-dependent
gravitational background in the vicinity of the collapsing matter. In
the asymptotic regime, the time dependent background can be eliminated
in favor of a dynamical boundary condition by an appropriate choice of
coordinates. This gives the mirror model description of black hole
emission. Locally, for free field propagation in radial modes, the
point $r=0$ is a perfectly reflecting boundary which behaves as a time
dependent brick wall, following an asymptotic  trajectory in tortoise
coordinates \begin{equation}\label{mirror}  r_*(r=0)\simeq
-t-Ae^{-t/2GM}+B\,. \end{equation}

In these models, the position of the infalling matter at late times
stays asymptotically at a fixed tortoise distance from the origin, and
provides a natural cutoff for the static Cauchy surfaces. At any
finite $t$, the spectrum of fields inside a large box is discrete,
becoming continuous only in the mathematical limit $t=\infty$, which
is totally unphysical because of the back reaction.  We can rewrite
Eq.\ (\ref{rshft}) in terms of the optical volume $\bar V$ outside
the infalling matter shell: \begin{equation}\label{sopv} S_{\rm
div}\simeq {d\;\Gamma(d/2)\;\zeta (d)\over \pi^{d/2}} {{\bar V}\over
\beta_H^{d-1}}\,. \end{equation} In two dimensions the optical volume
diverges linearly with the tortoise coordinate (logarithmically in
proper distance),  whereas in four dimensions
\begin{equation}\label{vdiv} \bar V_t\sim A_H GMe^{- r_*/2GM}\sim
(GM)^3 e^{t/2GM}\,. \end{equation}

If we want to regard Eq.\ (\ref{sopv}) as the geometric entropy
outside the infalling matter we must get rid of the boundary
divergence at the position of the outer shell. This can be done
following Ref.\ \cite{hlw}, by subtracting the geometric entropy in
the vacuum (the mirror remaining stationary). The result should be an
extensive entropy with respect to the optical volume (this was
explicitly checked in two dimensions in Ref.\ \cite{hlw}, and it is
very plausible in four dimensions as well). In any case, as the
tortoise position of the infalling matter recedes to $r_*\rightarrow
-\infty$, the optical volume diverges exponentially and we find the
divergence of 't~Hooft. Notice that in Eq.\ (\ref{vdiv}) the Newton
constant is the one entering in the mirror trajectory, i.e., the
renormalized $G$.

Therefore, if we regularize an eternal black hole by a physical
collapsing star, the continuous spectrum disease becomes an artifact
of the time slicing used inside the collapsing star, or else it
corresponds to the infinite entropy production at $t=\infty$.

The entropy source in these models is formally the mirror itself,
although a more accurate interpretation would be that the
time-dependent state of the infalling matter and gravitational fields
decays with a thermal cross section. In this sense, the difficulties
in locating the proper degrees of freedom of black hole entropy are
naturally due to the classical treatment of the radiation source.

\section{Discussion}
We have discussed several aspects, both
technical and conceptual, of the black hole entropy problem.  In
section \ref{thermal} we have shown that classical `intrinsic'
entropy makes sense in low energy effective theory even in a
non-equilibrium situation. The fact that it appears as a classical
object could be due to the use of stationary saddle points to
approximate the path-integral. After all, in the sphaleron
interpretation of black hole nucleation out of hot flat space one is
talking about a classical process of excitation {\it over} the
barrier, i.e., the nucleated black holes are formed by physical
collapse of graviton `matter'. But, of course, there are no temporal
Killing vectors inside the collapsing matter, even asymptotically.
The low energy theorem of Susskind and Uglum can be applied to this
situation provided a covariant regularization procedure is used
throughout. We also found  it useful to distinguish between
ultraviolet divergences in determinants of operators with discrete
spectrum, from others with continuous spectrum, such as the ones
appearing in the brick wall model. Then we have argued in favor of
fully covariant path integral prescriptions
 (leading to operators with discrete spectrum before ultraviolet
regularization) in systematic discussions of entropy renormalization.

Some recent proposals for the solution of the black hole entropy
problem and the information puzzle involve, in one way or another, a
breakdown of low energy effective field theory in the vicinity of the
horizon, even for big black holes and early stages of the evaporation
process. Since the discussions of black hole entropy renormalization
(particularly those based on Euclidean methods) assume the validity
of low energy field theory, this question becomes very relevant to
the matters discussed in this paper. Therefore, we would like to end
with some speculations on the related question of a low energy
description of the microscopic degrees of freedom responsible for
$S_{{\rm BH}}$.

{}From the point of view of mirror models, one would associate the
quantum degrees of freedom of black hole entropy with the radiation
source: the infalling matter and corresponding {\it time-dependent}
gravitational field. The problem, of course, is that this Hilbert
space has dimension $\sim A_H^{3/2}/\ell_{P\ell}^3$, instead of the
required $A_H/\ell_{P\ell}^2$. Here is where exotic quantum gravity
physics, such as the `holographic' phase \cite{th2,sho}, seems
unavoidable.

Actually, there is a natural notion of black hole entropy, closely
related to the phenomenological derivation of 't~Hooft given in
section \ref{intro}, which avoids explicit input from Planck-scale
physics. It is based on the idea that a black hole radiates not
because it is thermally excited in some way, but just because its
cross section for decay happens to be thermal.

In Hawking's approximation one computes the decay rate by scattering
the asymptotic vacuum off the time dependent {\it classical}
gravitational field. In a full quantum treatment the condition that
the external field approach is a sensible approximation can be
formalized by taking a coherent state for the infalling matter state
(and the induced graviton condensate). By a coherent state we mean a
minimum spread wave packet or, a state in which expectation values of
operators are given as classical functions of the expectation values
of the coordinates and momenta in the regularized theory (with a
cutoff in place). One would then work in a Hilbert space of the form
\begin{equation}\label{hcoh} {\cal H}_{\rm Haw}={\cal H}_{\rm
coherent}\otimes  {\cal H}_{\rm rad}\,, \end{equation} where states
are approximated by the product of a coherent time-dependent
infalling state $|\Psi_{\rm coh}(t)\rangle$ and a dilute radiation
state $|\omega_1\dots\omega_n\rangle$ of $n$ Hawking quanta. The
interaction Hamiltonian in the extreme coherent approximation would
induce an effective time-dependent background field potential for
Hawking quanta: \begin{eqnarray}\label{effv} &&\langle\Psi_{\rm
coh}(t)|\otimes\langle \omega_1\dots\omega_n| H_{\rm int}|\Psi_{\rm
coh}(t)\rangle\otimes|0\rangle \nonumber\\ &\simeq&
\langle\omega_1\dots\omega_n| V_{\rm eff} (g_{\mu\nu}(t))|0\rangle\,.
\end{eqnarray}

This yields Hawking's analysis. However, quantum back reaction changes
this picture, since after each radiative transition the initial
coherent state slightly decoheres. One has
\begin{equation}\label{decoh} |\Psi_{\rm coh}(t)\rangle\otimes
|0\rangle \rightarrow |\Psi_\omega (t)\rangle\otimes|\omega\rangle\,,
\end{equation} where $|\Psi_\omega\rangle$ is at best quasi-coherent,
and is distributed depending on $\omega$ (i.e., it is entangled with
$|\omega\rangle$). If $|\Psi_{\rm coh}(t)\rangle$ has mass $M$, then
$\hat M|\Psi_\omega(t)\rangle = (M-\omega)|\Psi_\omega(t)\rangle$.

It is clear that most of the $A_H^{3/2}/\ell_{P\ell}^3$ states of the
infalling Hilbert space are not quasi-coherent and, therefore, if
excited they do not decay thermally at all. For example, if a
super-Planckian Hawking quantum is generated with $\omega\sim M/2$,
then, obviously, the entangled states $|\Psi_{M/2}(t)\rangle$ must be
very far from being coherent. Of course, during the first stages of
the evaporation process we know that, as long as Hawking's
computation is accurate, most quanta have $\omega \sim (GM)^{-1}$
and, since $(GM)^{-1} \ll M$, then all the states $|\Psi_{1/GM}
(t)\rangle$ should be quasi-coherent. How many of these states are
there? This is a difficult computation to do, but one can estimate
their number by counting the number of ways to extract independent
subsystems of energy $(GM)^{-1}$ from a system of energy $M$:
\begin{equation} \label{cohonuda} {\rm dim}\left\{| \Psi_{1/GM} (t)
\rangle \right\} \sim {M\over \langle \omega \rangle} \sim G M^2 \sim
S_{{\rm BH}}\,. \end{equation} That is, the correct order of
magnitude. We think that this notion of quasi-coherence as a basis
for black hole entropy is the closest to the spirit of the
phenomenological derivation of the entropy based on Eq.\
(\ref{thoofts}) and, most importantly, it does not necessarily rely
on unknown quantum gravity effects, which could pollute Hawking's
calculation even in the earliest stages of the evaporation process.

In any case, if important deviations from thermality should occur from
the beginning, variants of this picture can be accomodated. For
example, if we consider a set of infalling states where the decay
cross section has a (not necessarily thermal) profile
\begin{equation}\label{profi} \Gamma_{\rm out}\sim A_H e^{-f(M,\delta
M)}\,, \end{equation} then, following the discussion in the
introduction, the entropy associated to this subset of the Hilbert
space is \begin{equation}\label{subset} S\sim\int dM {\partial f\over
\partial(\delta M)}(M,0)\,, \end{equation} which does not necessarily
scale as the  horizon area.

It would be very interesting to further study these notions in
simplified models.

It is amusing to speculate what this picture implies for the late
stages of the evaporation process. With the definition
(\ref{cohonuda}), $S_{{\rm BH}}$ is clearly decreasing in time,
because the quasi-classical infalling state progressively decoheres.
It is clear that, after a number of soft emissions of order
$GM^2$---so that the remaining mass is, say $M/2$---then the
infalling state is very poorly approximated by an external classical
field. Therefore,  further decay will not proceed with a thermal
cross section; it seems that the infalling matter can become `fuzzy'
still at macroscopic masses, thus spoiling Hawking's prediction long
before higher derivative gravity counterterms become important. In
contrast with other scenarios \cite{th2,ver},  this would be a purely
`soft' resolution of the information puzzle. Of course, under these
conditions the `operational' version of the Equivalence Principle is
violated: any infalling observer trying to experience a smooth
transition through the horizon would have lost its classical
properties in a much earlier stage.

\section*{Acknowledgements}

It is a pleasure to thank C.\ Callan, D.\ Gross, F.\ Larsen, and A.\
Peet for useful discussions. The work of JLFB was supported by NSF
Grant No.\ PHY90-21984,  while RE was partially supported by a FPI
grant  from MEC (Spain) and projects UPV 063.320-EB119-92 and CICYT
AEN93-0435. RE would also like to thank the High Energy Physics Group
at Princeton University for hospitality during the last stages of
this work.

\appendix
\section*{Appendix A}\label{ap}
 In this appendix we  check explicitly
in Rindler space the equality of the Green's functions of the
operators $\widehat{\cal L}$ and ${\cal L}$, when  both are  computed
with no boundary condition at the horizon.

The Green's function of $\widehat{\cal L}$ can be written as [see
Eq.\ (\ref{greenf}] \begin{eqnarray}\label{ghat} \widehat
G(x,x')&=&\langle x|\widehat{\cal L}^{-1}|x'\rangle \nonumber\\
&=&\sum_{n, \omega}\left({4\pi^2 n^2\over\beta^2}
+\omega^2\right)^{-1} e^{2\pi i n \Delta\theta/\beta} \psi_\omega^*
({\bf x}) \psi_\omega  ({\bf x'})\,, \end{eqnarray}      where
$\psi_\omega({\bf x})$ are eigenfunctions of Eq.\ (\ref{eigenfreq})
for the particular case of Rindler space.  Clearly, this yields the
solution to Eq.\ (\ref{greeneq2}) in the text.

The metric of $d$-dimensional Euclidean Rindler space is
\begin{equation}\label{rindmet}
ds^2=\xi^2d\theta^2+d\xi^2+\sum_{j=1}^{d-2}dx_j^2\,, \end{equation}
where $\theta$ is the Euclidean time, $\xi$ is the coordinate that
labels constant acceleration trajectories, and $x_j$ are flat
transverse coordinates. In these coordinates, and after separation of
transverse space variables, Eq.\ (\ref{eigenfreq}) takes the form of a
Bessel equation. As stressed in the text, an  important feature here
is that in the absence of a cutoff for small $\xi$, the spectrum of
frequencies $\omega$ is continuous.

The following expression for the Green's function of a massive scalar
can be readily obtained:  \begin{eqnarray}\label{greb} \widehat
G^{\beta}(x,x')&=&{1\over \pi^2}\int \prod_{j=1}^{d-2} {dp_j\over
2\pi} e^{ip_j \Delta x_j}\int_0^\infty d\omega \sinh
\pi\omega\nonumber\\ &\times& K_{i\omega}(\mu\xi)K_{i\omega}
(\mu\xi')\sum_{k=-\infty}^{+\infty}
 e^{-\omega|\Delta\theta+k\beta|}  \,,  \end{eqnarray}     where
$\mu^2\equiv m^2+\sum_jp_j^2$, and $K_{i\omega}(\mu\xi)$ are modified
Bessel functions.

Now, the sum over $k$ can be performed as \begin{equation}
\sum_{k=-\infty}^{+\infty}e^{-\omega|\Delta\theta+k\beta|} =
{\beta\over 2\pi} {\cosh\omega(\Delta\theta-\beta/2)\over\sinh
\beta\omega/2}\,.  \end{equation} We will be interested precisely in
$\beta=\beta_H=2\pi$.

Transverse momenta can also be integrated (details on similar
manipulations   can be found in Ref.\ \cite{car}):
\begin{eqnarray}\label{momint} &&\int\prod_{j=1}^{d-2} {dp_j\over
2\pi} e^{ip_j\Delta x_j} K_{i\omega}(\mu\xi)
K_{i\omega}(\mu\xi')\nonumber\\  &=&{1\over 2}\int_{-\infty}^{\infty}
d\lambda \; e^{i\omega\lambda} \left({m\over
2\pi\gamma}\right)^{d-2\over 2} K_{d-2\over 2}(m\gamma)\,,
\end{eqnarray}   with $\gamma^2(\lambda)\equiv
\xi^2+\xi'^2+2\xi\xi'\cosh\lambda+ \sum_j(\Delta x_j)^2$.

Therefore, \begin{eqnarray}  \widehat G^{2\pi}(x,x')&=& {1\over
2\pi^2}\int_0^\infty d\omega \cosh\omega
(\Delta\theta-\pi)\int_{-\infty}^\infty d\lambda\; e^{i\omega\lambda}
\nonumber\\  &\times& \bigl({m\over 2\pi\gamma}\bigr)^{d-2\over 2}
K_{d-2\over 2}(m\gamma)\,. \end{eqnarray}  At this moment we want to
interchange the order of integrations.  Convergence then requires
${\rm Im}\lambda>|\Delta\theta-\pi|$, so that after integrating
$\omega$ we find \begin{eqnarray}   \widehat G^{2\pi}(x,x')&=&
{i\over 2\pi^2}\int_C {\lambda d\lambda\over
\lambda^2+(\Delta\theta-\pi)^2} \nonumber\\ & \times&\bigl({m\over
2\pi\gamma}\bigr)^{ d-2\over 2} K_{d-2\over 2} (m\gamma)\,,
\end{eqnarray} where the contour $C$ runs from $-\infty$ to $+\infty$
passing above the pole at $\lambda=i|\Delta\theta-\pi|$. We can split
the integration contour into a straight line along the real axis and a
clockwise contour encircling the pole. The former contribution
vanishes by antisymmetry of the integrand, whereas the latter yields
\begin{equation}\label{gedospi} \widehat G^{2\pi}(x,x')={1\over
2\pi}\biggl({m\over 2\pi\sqrt{2\sigma}}\biggr)^{d-2\over 2}
K_{d-2\over 2}(m\sqrt{2\sigma})\,, \end{equation} where
$2\sigma\equiv \xi^2+\xi'^2-2\xi\xi' \cos\Delta\theta +\sum_j (\Delta
x_j)^2$    is the geodesic separation between the points $x, x'$ as
written in Rindler coordinates.  Then Eq.\ (\ref{gedospi}) is
precisely the Euclidean, zero-temperature, Green's function in
Minkowski space, i.e., the Hartle-Hawking Green's function, with no
boundary condition placed at the horizon. It must be noted  that Eq.\
(\ref{gedospi}) admits an expansion into Bessel functions of integer
order, corresponding to the standard solution of the Laplacian ${\cal
L}$ in ${\rm Disk}\times R^{d-2}$,
 regular at the origin and  with  discrete frequency spectrum.

\section*{Appendix B}\label{ap2d} In this appendix we briefly study the
possible
thermal instabilities of the linear dilaton vacuum of two dimensional
dilaton gravity, along the lines of the four dimensional analysis of
Ref.\ \cite{gpy}. This is an interesting exercise because Euclidean
gravity is on a much firmer ground in two dimensions and there is a
chance that all manipulations have a meaning in Lorentzian signature.
For example, string theory in the light cone and the Euclidean
covariant approach provides an example of such an equivalence.

The Euclidean action of two dimensional dilaton gravity is
\cite{cghs}  \begin{eqnarray}\label{action2d} I&=&-{1\over
2}\int_{\cal M}e^{-2\varphi}[R+4(\nabla\varphi)^2+4\lambda^2]
\nonumber\\ &-&\oint_{\partial{\cal M}}e^{-2\varphi}K +C_\infty\,,
\end{eqnarray} where $C_\infty$ is determined for our purposes by
requiring that, on a Hamiltonian thermal manifold, $I({\rm
cylinder})=\beta M_{\rm ADM}$.

In the conformal gauge
$g_{\alpha\beta}=e^{2\rho}\delta_{\alpha\beta}$ we have
\begin{eqnarray}\label{iconfg} I&=&-{1\over 2}\int_{\rm Disk}\left[-2
e^{-2\varphi} \partial^2(\rho -\varphi) +4\lambda^2
e^{2(\rho-\varphi)}\right] \nonumber\\ &-&\oint_{\partial{\cal M}}
e^{-2\varphi}K+\lambda \oint_\infty e^{-2\varphi}\,. \end{eqnarray}

The (Euclidean) classical black holes are parametrized by the mass $M$
\begin{equation}\label{twodbh} ds^2={d\sigma^2+d\theta^2\over
1+e^{-2\lambda\sigma}M/\lambda } \end{equation} and a dilaton
$\varphi=-{1\over 2}\log[M/\lambda-\exp(2\lambda\sigma)]$ where
$\sigma$ is a tortoise coordinate (the horizon is at
$\sigma=-\infty$). The solution with $M=0$  is the linear dilaton
vacuum: $g_{\alpha\beta}=\delta_{\alpha\beta}$,
$\varphi=-\lambda\sigma$, which becomes strongly coupled at left
infinity. In this case, unlike in  four dimensional black holes, the
Hawking temperature is unrelated to the mass and only depends on the
cosmological constant, $T_H=\lambda/2\pi$. This is an important
difference, since it implies that all black holes have the same
temperature and that the phenomenological entropy is proportional to
the mass $S_{{\rm BH}}=2\pi M/\lambda$. The classical suppression
factor for black hole nucleation vanishes in this case as
\begin{equation}\label{iclas} I_{\rm cl}(M)=\beta
M-\oint_{\infty}e^{-2\varphi}K=\beta M -\beta M =0\,. \end{equation}
Also, if we set $\beta\neq 2\pi/\lambda$, thus going off-shell,
\begin{eqnarray}\label{ioffs} I_{\rm cl}(M,\lambda,\beta)&=&\beta M
-2\pi e^{-2\varphi_H}\chi ({\rm Disk})  \nonumber\\ &=&\beta M -{2\pi
M\over \lambda}\,, \end{eqnarray}  where $\varphi_H$ is the value of
$\varphi$ at the horizon and $\chi$ is the Euler-Poincar\'e
characteristic.  Therefore the conical singularity method yields the
right answer for the entropy, as well as the classical method, Eq.\
(\ref{entdef}),  since at the critical temperature
\begin{equation}\label{scrit} S_{{\rm BH}}=\beta M -I_{\rm cl}=\beta
M={2\pi M\over \lambda}\,. \end{equation}

Now, the one-loop computation of the free energy around a particular
instanton is similar to that in Ref.\ \cite{gpy}.  Here we have a
renormalizable theory, but the position-dependent coupling
$g_s=e^\varphi$ makes it very difficult the non-perturbative analysis
of the path integral.

At a perturbative level there is a potential instability coming from
the fact that the dilaton field has the wrong metric. In this respect,
it plays a role similar to the conformal factor of the metric in four
dimensional gravity, and should not be considered as a physical
excitation. In fact, pure two dimensional dilaton gravity has no
propagating degrees of freedom. This is readily seen in the Lorentzian
path integral with the action (\ref{iconfg}). The functional
integration over $\varphi$ induces the condition that $\rho-\varphi$
be harmonic, so we can choose a (Kruskal) gauge in which
$\rho=\varphi$. If we want to mantain this in the Euclidean path
integral, we must integrate $\varphi=\varphi_{\rm cl}+ i\delta\varphi$
over the imaginary axis, and this produces a functional $\delta$
function $\prod_x \delta (-\nabla^2 (\rho-\varphi_{\rm cl}))={\rm
det}^{-1}(-\nabla^2 ) \prod_x \delta(\rho -\varphi_{\rm cl})$. This
enforces $\rho=\varphi_{\rm cl}$ and the determinant is cancelled by
the ghost determinant.

The analysis goes through if one adds appropriate counterterms to
take care of the one-loop conformal anomalies. Here one finds many
variants of the same model. For example, the one-loop action studied
in Refs.\ \cite{svv,chv} is constructed such that the manipulations
above make sense with $\exp(-2\varphi)$ replaced by $\Omega\equiv
\exp(-2\varphi) + N\varphi/24$. In general, the effective action must
preserve conformal invariance, and by means of non-linear field
redefinitions from $\rho$, $\varphi$ to new fields $X$, $Y$, one can
map the model to an open string theory \cite{cab}:
\begin{eqnarray}\label{calbil} I&=&-{1\over 2}\int [-(\partial X)^2+
(\partial Y)^2 +4\lambda^2 e^{C(X-Y)}]  \nonumber\\&+&I_{\rm
boundary} +I^{(N)}_{\rm matter}\,. \end{eqnarray} So we see that $Y$
works like a target time. In Lorentzian quantization one must cancel
$X$ against $Y$, leaving the $N$ `transverse' matter excitations. In
Euclidean quantization one must rotate $Y\rightarrow iY$ as well, so
that ${\rm det}^{-1}(-\nabla^2)$ from the $X$, $Y$ integrals cancels
against the ghost determinant.

As a result, for $N$ scalar matter fields the perturbative partition
function is proportional to ${\rm det}^{-N/2}(-\nabla^2)$, which is
positive definite. No imaginary part of the free energy is generated,
and consequently there is no black hole nucleation. In addition, the
absence of propagating gravitons rules out any possible infrared
Jeans instability.

This absence of tunneling barrier is compatible with the classical
canonical thermodynamical analysis. The free energy for the combined
system of two phases is (we neglect the boundary free energy)
\begin{equation}\label{canon} {} F=F_{\rm rad}+F_{{\rm
BH}}=-{\pi\over 6} NLT + M\left( 1-{2\pi\over \lambda} T\right)\,.
\end{equation} At the critical temperature $T_H=\lambda/2\pi$ there
is a flat direction in $M$, and the canonical ensemble makes sense
for two dimensional black holes, at least within perturbation theory.

It is also interesting to analyze the classical microcanonical
ensemble, where one maximizes the combined entropy $S=(\pi/3) NLT+
2\pi M/\lambda$ at fixed total energy $E=(\pi/6) NLT^2 + M$.   The
result in this case is very different from that in four dimensions
\cite{gip}. If the energy density $\varepsilon =E/L$ is less than a
critical value $\varepsilon_c =\lambda^2 N/(24 \pi)$ then we have
pure radiation with temperature $T=\sqrt{6\varepsilon/\pi N}$. Above
this energy the temperature remains constant $T_H=\lambda/2\pi$ and
the mass of the black hole grows linearly as $M=E-\varepsilon_c V$.

Regarding the one-loop divergences of the entropy, it is well known
\cite{suu} that the logarithmic divergence $S_{\rm div}=N/6
\log\epsilon^{-1}$ from $N$ matter fields contributes an infinite
additive constant to $S$ and cannot be renormalized in $\lambda$. In
this respect, two dimensional black holes follow a pattern different
from their four dimensional counterparts.

\end{document}